\documentclass{aa}  

\usepackage{graphicx}
\usepackage{txfonts}
\usepackage{graphicx}
\usepackage{mathrsfs}
\usepackage{url}
\begin{document}

   \title{Fast molecular jet from L1157-mm
   \thanks{Based on IRAM 30m Telescope observations.
   IRAM is supported by INSU/CNRS (France), MPG (Germany), and IGN (Spain).}
   }

   \author{M. Tafalla
          \inst{1}
          \and
          R. Bachiller
          \inst{1}
          \and
          B. Lefloch
          \inst{2}
          \and
          N. Rodr\'{\i}guez-Fern\'andez
          \inst{3}
          \and
          C. Codella
          \inst{4}
          \and
          A. L\'opez-Sepulcre
          \inst{2,5}
          \and
          L. Podio
          \inst{4}
          }

   \institute{Observatorio Astron\'omico Nacional (IGN), Alfonso XII 3,
              E-28014 Madrid, Spain\\
              \email{m.tafalla@oan.es}
              \and
       Univ. Grenoble Alpes, IPAG, F-38000 Grenoble, France\\
       CNRS, IPAG, F-38000 Grenoble, France
        \and
        CESBIO (UMR 5126), Obs. de Midi-Pyr\'en\'ees (CNES, CNRS, 
        Univ. Paul Sabatier, IRD), 31401 Toulouse, France
        \and
        INAF, Osservatorio Astrofisico di Arcetri, Largo E. Fermi 5, 
        50125 Firenze, Italy
        \and
        Department of Physics, The University of Tokyo, Bunkyo-ku, 
        Tokyo 113-0033, Japan
             }

   \date{}

  \abstract
   {L1157-mm powers a molecular outflow that is well-known for its
      shock-induced chemical activity in several hot-spots.}
   {We have studied the molecular emission toward L1157-mm 
   searching for a jet component responsible for these spots.}
   {We used the IRAM 30m telescope to observe the vicinity of 
   L1157-mm in several lines of SiO.} 
   {The SiO(5--4) and SiO(6--5) spectra toward L1157-mm present
   blue and red detached components about 45~km~s$^{-1}$ away from
   the ambient cloud. 
   These extremely high-velocity (EHV) components are 
   similar to those found in the 
   L1448 and IRAS 04166+2706 outflows and probably arise from 
   a molecular jet driven by L1157-mm.
   Observations of off-center positions indicate that the jet is unresolved
   in SiO(5--4)  (<$11''$).}
   {The EHV jet seen in SiO probably excites L1157-B1
   and the other
   chemically active spots of the L1157 outflow.
   }


   \maketitle
%

\section{Introduction}

The \object{L1157} outflow 
was first identified by its unusual heating of the ambient gas
toward the so-called B1 position; this is
suggestive of a
strong shock interaction \citep{ume92}.
Many observations have confirmed this result
and showed that molecular abundances in
L1157-B1 and other hot-spots 
are enhanced by orders of magnitude
in a manner consistent with
models of shock chemistry 
\citep{mik92,bac97}.
The richness and intensity of the molecular
spectrum from L1157-B1 has made this position a testbed
for shock-chemistry models and a favored target
for molecular surveys \citep{cod10,lef10,yam12}.

The spectra from L1157 (like those from other outflows)
present broad wings. This means that most
of the emitting gas moves with the lowest outflow velocities
and that only a small fraction of the gas reaches
the highest speeds. This
distribution indicates that the emitting gas 
consists of accelerated ambient material and 
that the accelerating agent
remains hidden.

A number of observations suggest that the exciting agent of the
L1157 hot-spots is a jet.
At high angular resolution, the emission from L1157-B1
resembles a bow shock whose vertex points away from the
exciting source \citep{taf95,gue96}, while high-excitation
lines trace the reverse shock \citep{ben12}.
In addition, the blue and red hot-spots 
occur in pairs located symmetrically from
the L1157-mm central source, and they seem to trace
a slowly precessing jet \citep{zha00,bac01}.
Despite these indications, however,
no direct detection of a jet component such as
that seen in the \object{L1448} and \object{IRAS 04166+2706}
outflows has been presented for L1157.
\citet{kri12} have recently identified a red ``bullet'' in the 
water spectrum of L1157-mm.
This feature could correspond to a true jet or to some other type of
shock near the protostar.
In this paper, we present 
SiO observations of the L1157-mm vicinity that convincingly show
that a molecular jet is at the base of the 
L1157 outflow.

\section{Observations}

Our main dataset is a line survey toward
L1157-mm carried out as part of the Astrochemical Surveys At IRAM (ASAI)
project (Lefloch \& Bachiller, in prep.).
This project covered
the 1, 2, and 3~mm frequency bands that are reachable with the IRAM 30m 
telescope\footnote{\url{http://www.oan.es/asai/}}. 
L1157-mm was observed in September 2012 and February
2013 using the EMIR receiver and 
the FTS spectrometer, which provided a frequency
resolution of 200~kHz.
Wobbler-switching by
$180''$ was used to ensure flat baselines, and each frequency
setting consisted of two tunings 50 MHz apart to 
separate lines coming from the upper and lower sidebands.

Follow-up SiO observations were carried out with the 30m
telescope in January-February and
March 2014 using a setting similar
to the ASAI project. They included integrations on 
L1448mm and IRAS 04166+2706 (IRAS 04166 hereafter).
Additionally, we here used 30m data from two 
deep CO(2--1) integrations carried out in February 2001, also
with similar settings.

The ASAI and 2014 data were compared and 
combined. The intensity of SiO(6--5) in these two datasets
differed by a factor of 1.5 of unclear origin, and
to avoid any bias, the data were averaged with equal weight. 
Standard
efficiency values were applied to convert the data 
into to the $T_{\mathrm{mb}}$ scale used in the paper.
A summary of the observations is presented in 
Tables~\ref{tbl_parm} and \ref{tbl_out}.

\section{Spectral evidence for a jet in L1157-mm}
\label{sect_spec}

\begin{figure}
\centering
\resizebox{7cm}{!}{\includegraphics{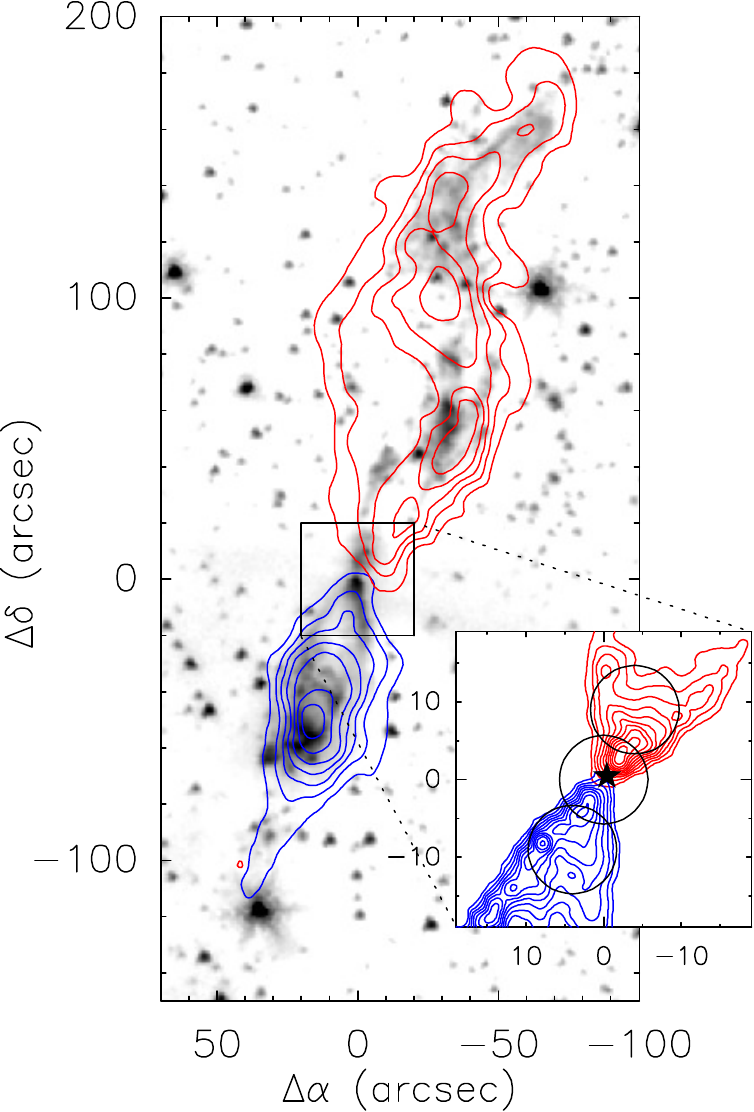}}
\caption{Large-scale view of the L1157 outflow. The main frame shows 
a superposition of the CO map from \citet{bac01} (contours) and the
3.6~$\mu$m Spitzer image of \citet{loo07}. The inset shows
the inner outflow according to the
interferometric CO map of \citet{joe07}. 
The star symbol represents L1157-mm and the three 
circles represent the positions and beam sizes of our SiO(5--4) 
observations.
\label{co_map}}
\end{figure}

Figure~\ref{co_map} illustrates the geometry of the
L1157 outflow with a superposition of 
a CO map (from \citealt{bac01}) and a
3.6~$\mu$m Spitzer image (from \citealt{loo07}). 
An inset shows a high-resolution CO map
toward the vicinity of the  young stellar object (YSO, adapted from \citealt{joe07}, see also \citealt{hul14}). 
This inner image reveals two opposed
conical shells with the protostar located at their apex.
The three circles mark
the positions of the SiO(5--4)
observations along the outflow discussed in Sect.~\ref{sect_extent}.

\begin{figure}
\centering
\resizebox{0.9\hsize}{!}{\includegraphics{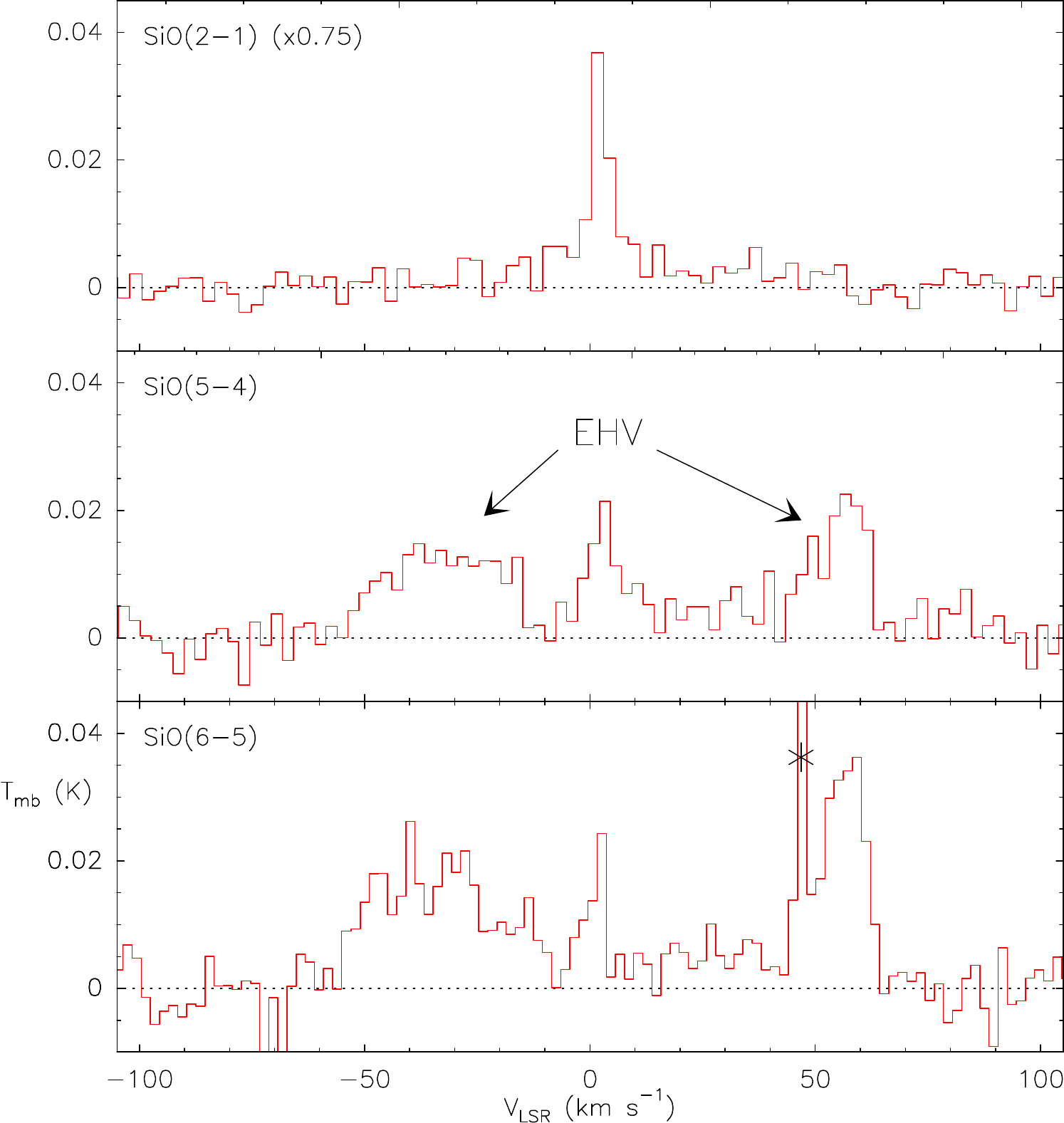}}
\caption{SiO spectra toward 
L1157mm (RA(J2000) = $20^{\mathrm h}39^{\mathrm m}06\fs3$,  
Dec(J2000) = $68\degr02'15\farcs8$).
The component near 2.7~km~s$^{-1}$, most prominent in
SiO(2--1), arises from accelerated ambient gas, while the
two components labeled ``EHV'' 
probably represent a molecular jet.
An asterisk in the SiO(6--5) spectrum marks the contribution of
ambient $c$-C$_3$H$_2$(5$_{32}$--4$_{41}$).
\label{sio_spec}}
\end{figure}

Figure~\ref{sio_spec} shows the
SiO(2--1), (5--4) and (6--5) spectra
toward L1157-mm 
(the noisier SiO(3--2) ASAI spectrum 
is not shown).
The SiO(2--1) emission in the top panel consists of
a narrow feature 
at the cloud velocity (2.7~km~s$^{-1}$, \citealt{bac01})
and low-intensity outflow wings.
In contrast, the SiO(5--4) and (6--5) spectra (middle and bottom panels)
show progressively weaker ambient features and increasingly more 
prominent blue and red detached components.

A comparison between the two ASAI spectra obtained with receiver
tunings separated by 50~MHz rules out contamination by lines in 
the image sideband as a possible origin of the detached components.
In addition, a
search in the JPL and CDMS spectroscopy databases \citep{pic98,mue01}
shows no alternative molecular lines at the corresponding
frequencies.
The detached components must therefore arise from 
extremely high-velocity (EHV) SiO emission 
in the outflow.
Previous evidence of EHV gas in L1157-mm
was provided by \citet{kri12}, who identified
a weak red ``bullet'' in the
H$_2$O($1_{10}$--$1_{01}$) line. 
Our new data confirm this H$_2$O detection, although the red
SiO EHV peak is 10~km~s$^{-1}$ faster and 1.5 times narrower
than the H$_2$O peak. 
More importantly, the new SiO data show 
an equivalent EHV feature at blue velocities.
While similar, the blue and red EHV components are not
symmetric. Both SiO(5--4) and SiO(6--5) spectra show
that the red EHV component is faster and narrower than the
blue component, which suggests that this reflects an intrinsic property
of the system.

\begin{figure}
\centering
\resizebox{0.9\hsize}{!}{\includegraphics{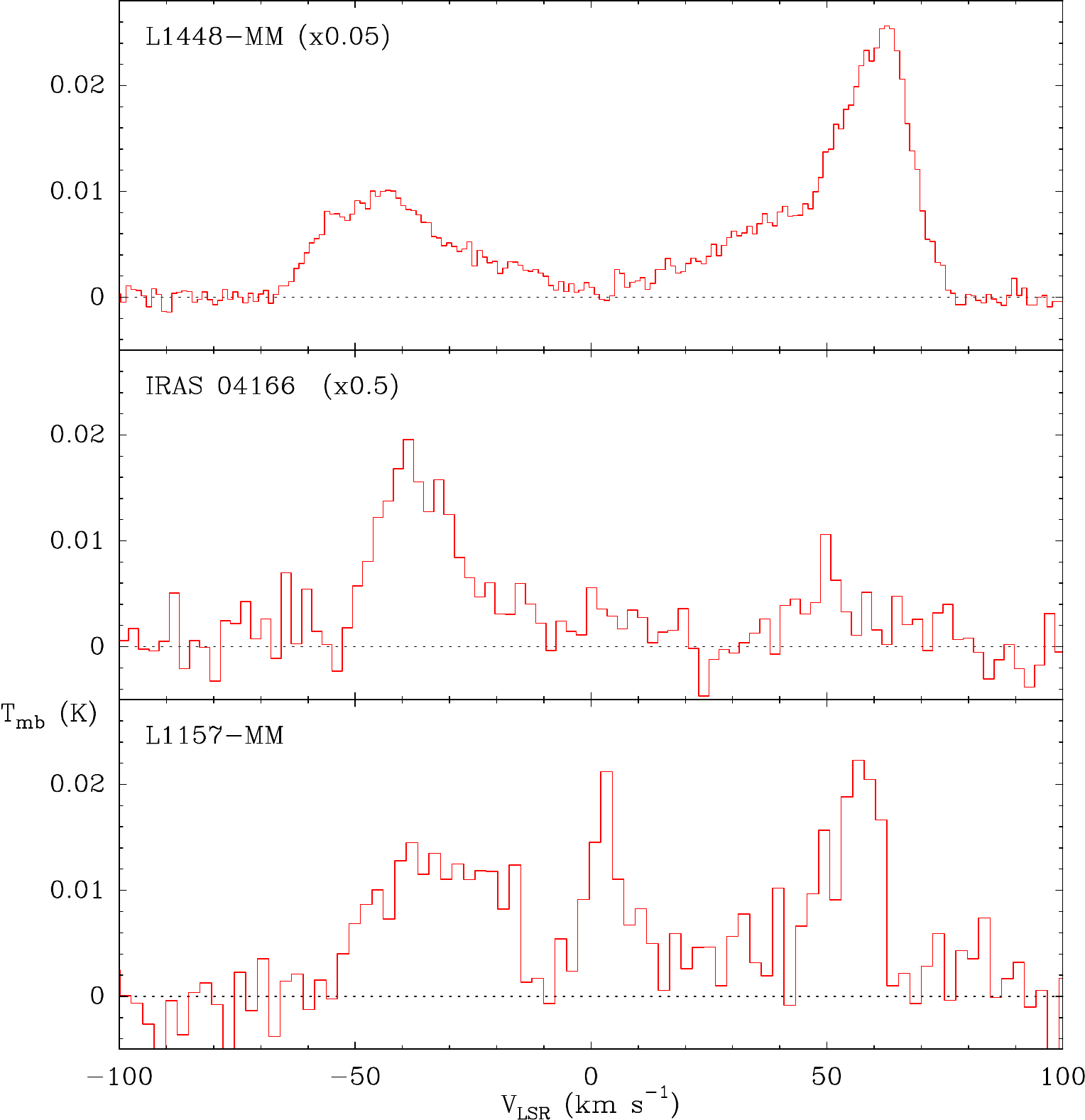}}
\caption{SiO(5-4) spectra toward the outflow
sources L1448-mm, IRAS 04166, and L1157-mm.
All sources present EHV components of similar
velocity but different intensity.
\label{sio_ehv}}
\end{figure}

Figure~\ref{sio_ehv} compares
the SiO(5--4) spectrum from L1157-mm with
equivalent spectra from L1448 and IRAS 04166.
It shows that
despite differences in intensity of up to a factor 
of 20, all EHV components 
have velocities of about 40-50~km~s$^{-1}$ with respect
to the ambient cloud, 
are broad ($\sim 10$~km~s$^{-1}$),
and often present asymmetric profiles with a flatter slope toward
ambient velocities and a steeper drop toward higher speeds.
The degree of symmetry between the blue and red EHV components 
varies over the sample,
with L1157-mm presenting the two components with
the most similar intensity (difference less than 40\%).

When the EHV emission of the L1448 and IRAS 04166 outflows
has been observed with 
high angular resolution, it has 
been found to arise from a distinct jet-like
component \citep{gui92,san09,hir10}.
This component moves
along the axis of the outflow cavity delineated by the 
lower-velocity gas and traces
a series of internal working surfaces in
a molecular jet \citep{san09}.
The finding of an EHV component toward L1157-mm
therefore reveals a fast
molecular jet in the L1157 outflow.

\section{Extent of the EHV emission}
\label{sect_extent}

\begin{figure}
\centering
\resizebox{0.9\hsize}{!}{\includegraphics{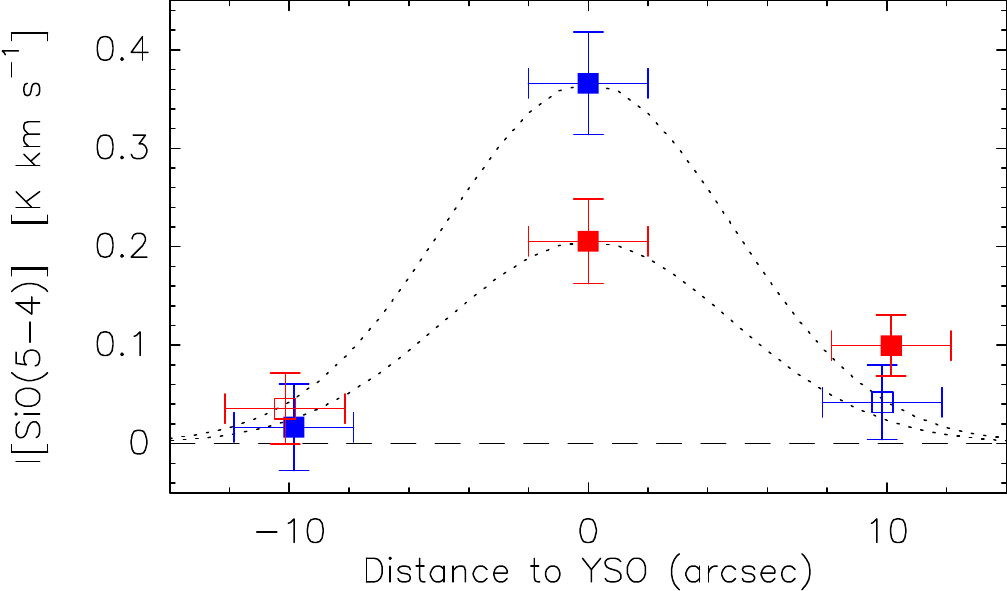}}
\caption{Integrated intensities of the SiO (5-4) EHV emission 
(both blue and red components) measured at three positions along the 
L1157 outflow axis. The negative and positive distances correspond to
the blue and red outflow lobes.
The dotted curves are Gaussians with the size of
the telescope beam at the SiO(5-4) frequency.
A nominal rms position uncertainty 
of $2''$ has been assumed for all positions.
\label{ehv_cut}}
\end{figure}

In the L1448 and IRAS 04166 outflows, the EHV emission 
is as extended as the low-velocity CO.
In L1157, the emission is very compact.
This is illustrated in Fig.~\ref{ehv_cut} with
an intensity cut along the axis of the outflow cavity. As a result
of the weak SiO lines, which require several hours of
integration, our observations of the EHV component in 
L1157 were limited to 
the three points indicated in Fig.~\ref{co_map},
whose positions are given in Table~\ref{tbl_out}.
Even with this limited dataset, it is possible to constrain
the emission size.

As Fig.~\ref{ehv_cut} shows, no significant detection of EHV
emission was obtained outside the central source.
To model this result, we compared
the data from each EHV component
with a Gaussian that has the FWHM of the telescope
beam at the SiO(5--4) frequency, which represents 
the expected intensity distribution for a point-like source
centered on L1157-mm.
The dotted lines in Fig.~\ref{ehv_cut} show that the blue
EHV emission, which is expected to extend to negative distances,
is consistent with being unresolved, while the
red emission (expected toward positive distances)
is at most marginally detected outside L1157-mm at 
a 3-sigma level. The size of the EHV emission must therefore be 
smaller than the $11''$ beam (a conversion to physical size
depends on the uncertain source distance of 440-250~pc,
\citealt{loo07}).

\section{SiO excitation, column density, and abundance}
\label{sect_excit}

\begin{figure}
\centering
\resizebox{0.9\hsize}{!}{\includegraphics{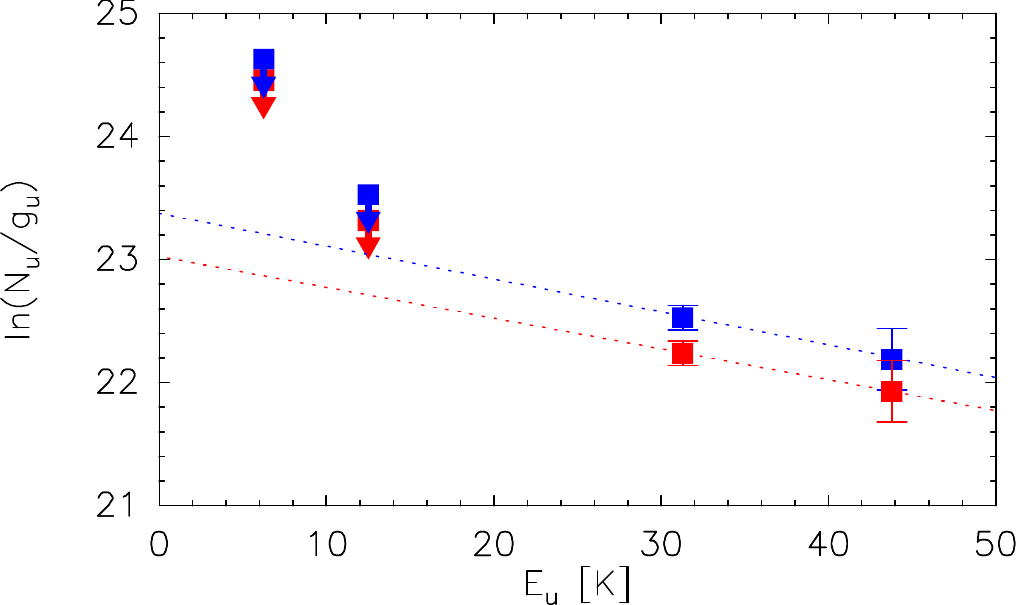}}
\caption{Population diagram of the EHV SiO toward L1157-mm. The blue and red
symbols represent blue and red EHV gas.
The dashed lines represent two LTE fits consistent with the
detections in SiO(5--4) and SiO(6--5) and with the 3-sigma upper limits
in SiO(2--1) and SiO(3--2) (downward pointing arrows).
See text for fit results.
\label{pop_rot}}
\end{figure}

After determining a limit to the size of the EHV emission,
we now use the intensity of the different lines 
to estimate the excitation temperature and column density 
of SiO in the EHV gas.
We assumed that the emitting region is unresolved
and scaled all line intensities to a common
size of $11''$ using point-source dilution factors.
These scaled intensities were then converted into upper-level 
column densities assuming  
optically thin emission. The resulting population diagram
is shown in Fig.~\ref{pop_rot}.

Since the EHV component is only detected in SiO(5--4) and SiO(6--5), 
we used only these two lines to estimate the SiO
excitation temperature and column density
assuming LTE conditions.
The error budget is dominated
by calibration uncertainties, which we assumed are
of 10\% for SiO(5--4) and of 25\% for SiO(6--5).
With these values, we estimate that the blue and red EHV components have 
a similar SiO excitation temperature of $40\pm 30$~K, 
while their SiO column densities are 
$(5\pm4) \times 10^{11}$ and $(4\pm3) \times 10^{11}$~cm$^{-2}$
.
As shown by the dashed lines in Fig.~\ref{pop_rot}, the
fits are 
consistent with the 
nondetections of
SiO(2--1) and SiO(3--2). Compared with 
similar estimates in the 
L1448 and IRAS 04166 outflows, the column densities in 
L1157-mm are lower by about an order of magnitude 
\citep{taf10}.

\begin{figure}
\centering
\resizebox{0.9\hsize}{!}{\includegraphics{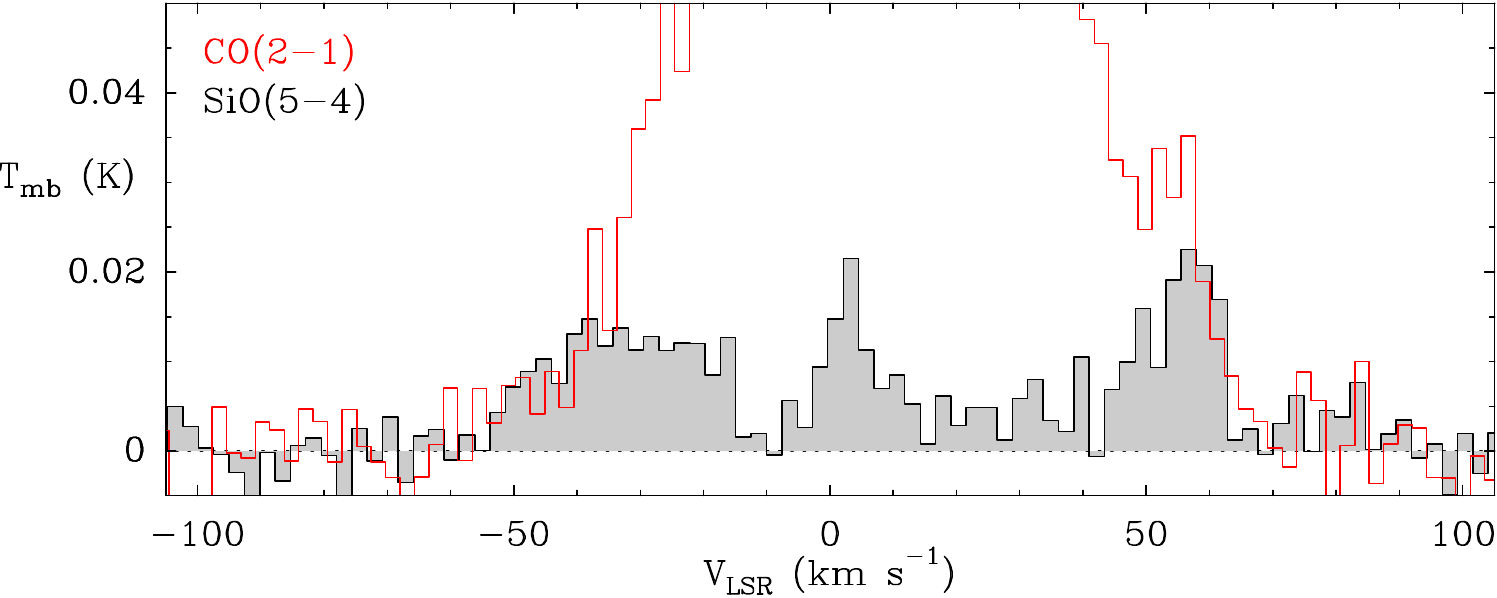}}
\caption{Spectra of CO(2--1) (red) and SiO(5-4) (gray) toward L1157-mm.
Note that the CO (2-1) line, which does not reveal distinct EHV emission, 
has similar velocity extents as the SiO (5-4) line with the same 
intensity level.
\label{co_sio}}
\end{figure}

To estimate the SiO abundance in the EHV component,
we used the intensity of CO(2--1), which was observed
by the ASAI project with an angular resolution similar to that
of SiO(5--4). The CO(2--1) spectrum is shown in Fig.~\ref{co_sio} and 
differs from that of SiO(5--4) in that it has no distinct EHV components. Instead, it has 
broad outflow wings that extend as far as the EHV range
(there is a hint of a red EHV peak at the noise level).
The different shape of the CO and SiO spectra is also seen
in L1448 and IRAS 04166 and
seems to be caused by the SiO emission being more selective of the
EHV component than the CO emission, which traces both the EHV gas and
the lower-velocity outflow. 
While it is not possible to separate the EHV and wing components in
the CO spectrum, the plot shows that the intensities of CO(2--1) and
SiO(5--4) are approximately equal at the velocities of the SiO EHV regime.

If we assume that the CO and SiO emissions are optically thin
and have the same excitation temperature of 40~K, the
observed CO(2--1)/SiO(5--4)
intensity ratio of 1 implies a column density
ratio of $\approx 500$. Assuming a CO abundance
of $8.5\times 10^{-5}$ \citep{fre82}, we
derive an SiO abundance of 
$\approx 1.7\times 10^{-7}$. This value agrees within a factor
of 2 with the abundance estimated by \citet{taf10}
in the EHV component of
the L1448 and IRAS 04166 outflows.

\section{Relation with L1157-B1}
\label{sect_b1}

\begin{figure}
\centering
\resizebox{0.9\hsize}{!}{\includegraphics{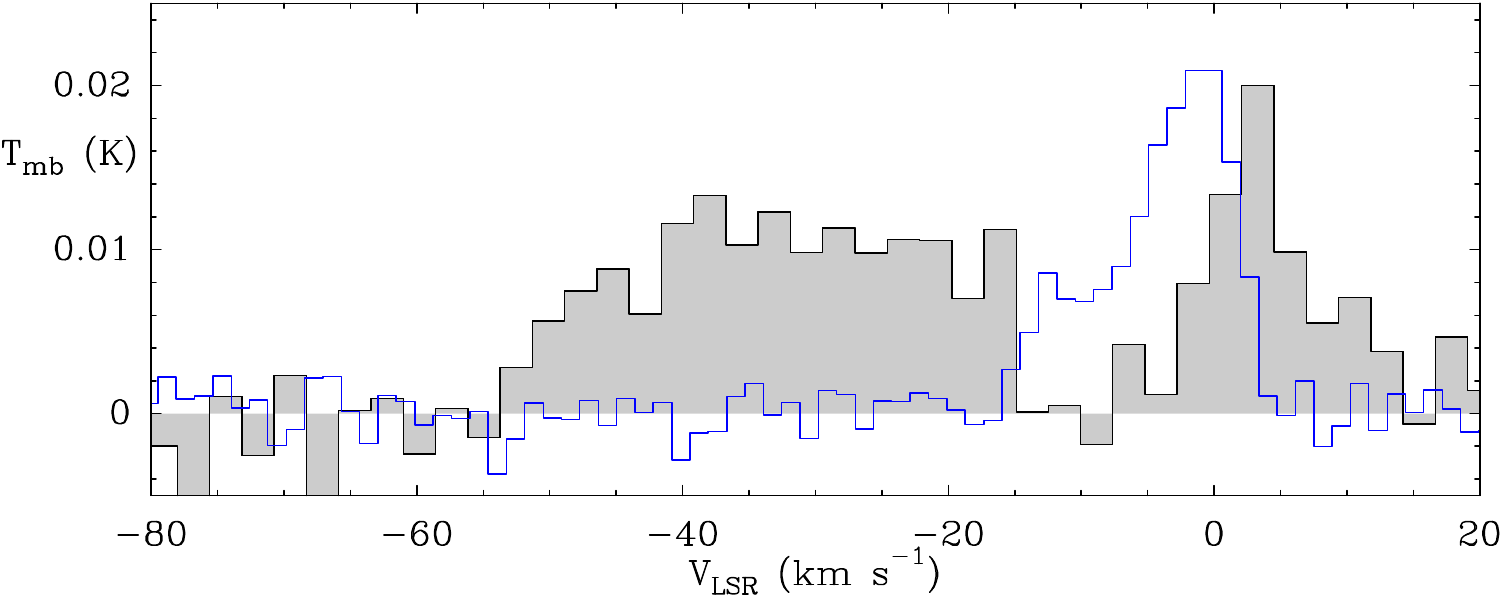}}
\caption{SiO(5--4) spectra toward L1157-mm (gray) and L1157-B1 (blue).
L1157-B1 spectrum scaled down by factor of 100.
\label{b1}}
\end{figure}

The EHV SiO jet is the
natural candidate for the exciting agent of L1157-B1 and the other outflow
hot-spots. 
To explore this possibility,
we compare in Fig.~\ref{b1} the SiO(5--4) emissions from
L1157-mm and L1157-B1 (from \citealt{bac01}).
As can be seen, the EHV component in L1157-mm is faster than
the main emission in the L1157-B1 spectrum,
as would be expected if the EHV gas is the driving agent.
(The EHV component reaches higher velocities than the 
fastest CO gas detected
in L1157-B1 to date, that is, 
the $g_1$ component of \citealt{lef12}, which is 
probably associated with the jet shock.)
To be more quantitative, we estimate from Fig.~\ref{b1} that
the blue EHV component has a mean velocity
of 35~km~s$^{-1}$ (from a Gaussian fit), and that
the main SiO emission in L1157-B1 reaches a velocity of about
18~km~s$^{-1}$.
If these two velocities correspond to the jet and bow shock 
components, their ratio approximately equals 2. 
A simple requirement of momentum conservation,
necessary for the jet to accelerate the shock,
implies that the velocity ratio equals $1+\sqrt{\rho/\rho_j}$, 
where $\rho$ and
$\rho_j$ are the ambient cloud and jet densities (see,
e.g., \citealt{mas93}).
Both the ambient and jet densities are very uncertain, but they
lie very probably in the vicinity of $10^5$~cm$^{-3}$, which is
typical of dense cores and molecular jets \citep{san09}. If so, 
their ratio should be close to 1, and $v_j/v_s$ should 
be $\approx 2$, as observed. 

\begin{figure}
\centering
\resizebox{0.9\hsize}{!}{\includegraphics{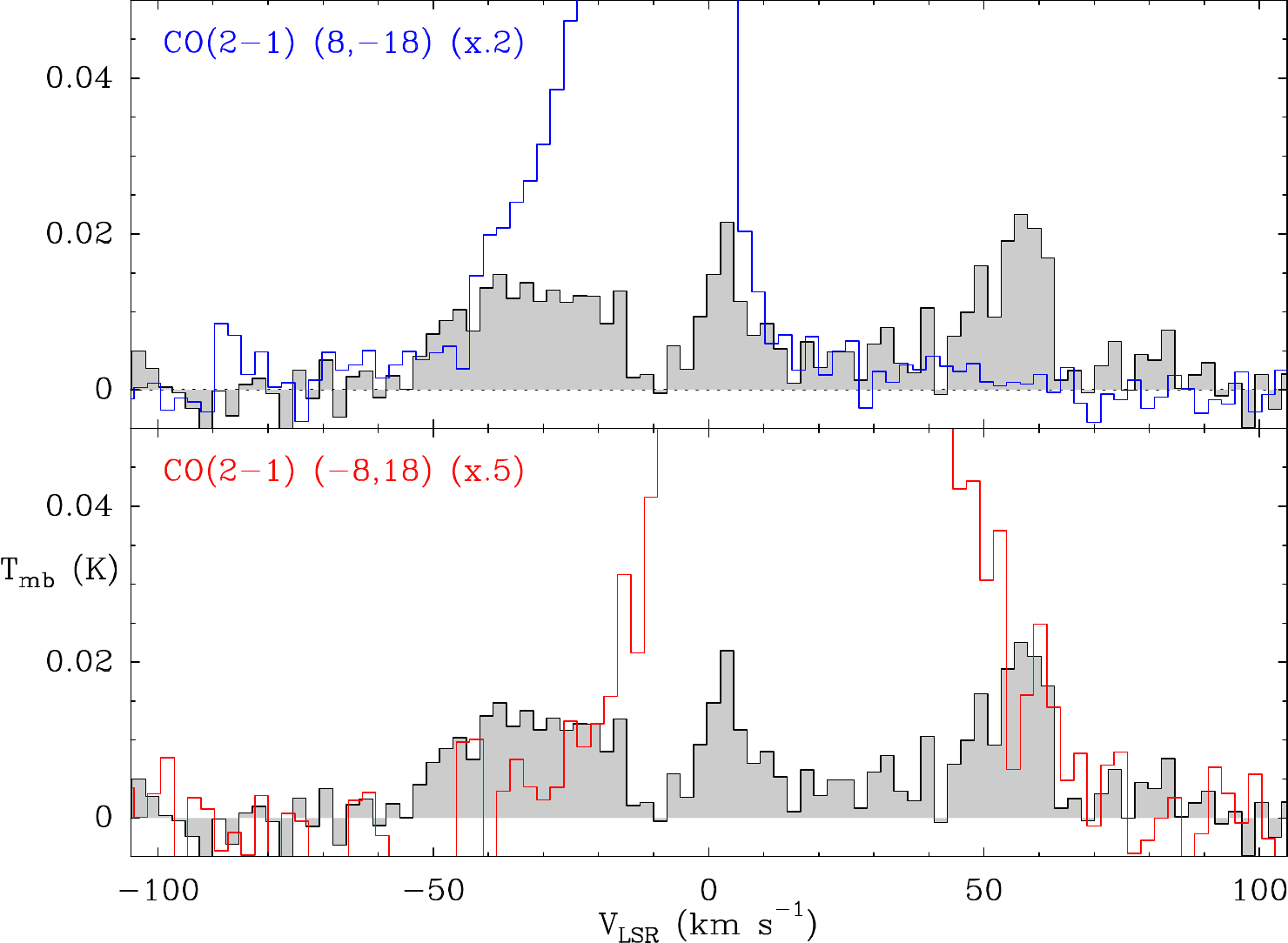}}
\caption{On-source SiO(5--4) spectrum versus CO(2--1) spectra $20''$ from 
the source.
In each panel, 
the SiO(5--4) spectrum toward L1157-mm is shown in gray and the 
CO(2--1) spectra at positions $20''$ away from L1157-mm are shown in color.
\label{co_off}}
\end{figure}

To be the accelerating agent,
in addition to having the correct velocity, 
the EHV gas needs to reach L1157-B1, which
is more than $60''$ away from L1157-mm. 
The EHV SiO emission
is detected only toward L1157-mm, but this likely results from a
combination of limited sensitivity and an intensity drop
of the SiO emission away from the central source 
similar to that seen in the L1448 and 
IRAS 04166 outflows \citep{san09,hir10}.
Thus, to confirm that the EHV gas reaches L1157-B1, we need to rely on
the CO emission, which is less selective of the 
EHV gas due to the broad wing component. 
Unfortunately, no large-scale deep CO observations
are available to follow the EHV emission between
L1157-mm and L1157-B1.
Our limited data do suggest that there is such an emission, however.
This is illustrated Fig. \ref{co_off} with CO(2--1) spectra 
toward two positions $20''$ away from L1157-mm in the direction of the 
expected jet. As can be seen,
both blue and red CO spectra extend in velocity as far as the SiO EHV
emission from L1157-mm, indicating that the EHV gas extends
farther than indicated by the SiO emission.
Thus, although further observations are needed to complete the
picture, our data suggest that the exciting jet of the 
L1157 outflow has finally been identified.

\begin{acknowledgements}
We thank the IRAM staff for assistance and the referee for a careful reading
of the manuscript.
We acknowledge support from projects
FIS2012-32096 and AYA2012-32032 of Spanish MINECO and from the MICINN program
CONSOLIDER INGENIO 2010, grant ``Molecular Astrophysics:
The Herschel and ALMA era - ASTROMOL'' (ref.: CSD2009-00038).
LP has received funding from the European Union Seventh Framework Programme 
(FP7/2007-2013) under grant agreement n. 267251.
This research has made use of NASA's Astrophysics Data System
Bibliographic Services.
\end{acknowledgements}


\Online

\begin{table}
\caption[]{Transitions observed toward L1157-mm.\label{tbl_parm}}
\centering
\begin{tabular}{lrrrr}
\hline\hline
\noalign{\smallskip}
Transition &   Frequency    &   FWHM     &  $\eta_{\mathrm mb}$ &
$\sigma(T_{\mathrm mb})$\tablefootmark{a} \\
         &      (GHz)      &  $('')$   & &  (mK) \\
\noalign{\smallskip}
\hline
\noalign{\smallskip}
SiO(2--1) & 86.8 & 28 & 0.81 & 3.8 \\
SiO(3--2) & 130.3 & 19 & 0.76 & 6.2 \\
SiO(5--4) & 217.1 & 11 & 0.62 & 3.3 \\
SiO(6--5) & 260.5 & 9 & 0.53 & 4.8 \\
\hline
\end{tabular}
\tablefoot{
\tablefoottext{a}{In a 1~km~s$^{-1}$ channel.}
}
\end{table}

\begin{table}
\caption[]{Additional observations toward the L1157 outflow.\label{tbl_out}}
\centering
\begin{tabular}{lrrrr}
\hline\hline
\noalign{\smallskip}
Transition &   Frequency    &   FWHM     & Offsets\tablefootmark{a} \\
         &      (GHz)      &  $('')$   & $('','')$ \\
\noalign{\smallskip}
\hline
\noalign{\smallskip}
SiO(5--4) & 217.1 & 11 & (-4,9), (4,-9) \\
CO(2--1) & 230.5 & 11 & (0,0), (-8,18), (8,-18) \\
\hline
\end{tabular}
\tablefoot{
\tablefoottext{a}{Measured from RA(J2000) = $20^{\mathrm h}39^{\mathrm m}06\fs3$,
Dec(J2000) = $68\degr02'15\farcs8$}
}
\end{table}

\end{document}